# Spaser in Coupled Metamaterials: The gain effect in a magnetic plasmon waveguide

Shuming Wang[1], Zhihong Zhu[1], Jingxiao Cao[1], Tao Li[1], Hui Liu[1,*], Shining Zhu[1], Xiang Zhang[2]

[1] Department of Physics, National Laboratory of Solid State Microstructures, Nanjing University, Nanjing 210093, People's Republic of China

[2] Nanoscale Science and Engineering Center, University of California, 5130 Etcheverry Hall, Berkeley, CA 94720-1740, USA

**Abstract:** Loss is a crucial problem in plasmonic integrated optical circuits and metamaterials. The Er, Yb codoped gain material is introduced into a magnetic plasmon waveguide composed of a chain of nanosandwiches in order to solve the loss problem in such subwavelength waveguide. The magnetic plasmon mode and a higher order mode are chosen as the signal and pump light to enhance the radiation and pump efficiencies. The signal light propagating in the waveguide is investigated with different $Er^{3+}$ doping concentration and signal decay time. It is shown that the gain effect can not only compensate the loss but also is able to amplify the signal, when exceeding certain threshold values of doping concentration and signal decay time.

The waveguides composed of chains of nobel metal nanoparticles have recently attracted widely attention in plasmonic integrated optical circuits. By using the coupled surface plasmon mode, which is able to confine the field tightly to the waveguide itself, one can get a small waveguide whose size even below the diffractive limit [1]. On the other hand, the subwavelength waveguide based on coupling of magnetic resonance between the SSRRs (single split ring resonator) has also been reported [2-4]. The collective magnetic resonance in the waveguide is called magnetic plasmon (MP). In such an MP waveguide, lower radiation loss and longer propagation length are obtained compared with the coupled electric resonance in nanoparticle chain. To overcome the frequency saturation of SSRR structure at about 100THz, the MP waveguide composed of metallic nanosandwiches is recently introduced and investigated, which is able to work in light frequency region and can present various usage, such as wavelength selective switch [5, 6]. However, there exists a fatal problem prevents these subwavelength waveguides from the realistic application that is the loss including the large scattering loss introduced by the micro-fabrication and the Ohmic loss of the metal component, especially at light frequency region. A promising method to compensate the loss in plasmonic systems is to combine the metallic structures with the gain materials [7-10]. In one of our recent work, a magnetic plasmon nanolaser is reported based on double resonance nanosandwich structures [11]. In this paper, we investigate the compensate effect in an MP waveguide combined with the ytterbium-erbium codoped gain material: Er:Yb:YCOB, and the lasing case is also found in this structure.

The geometry of the subwavelength MP waveguide is shown in Fig. 1(a). The nanosandwich is composed of two metallic rectangular slabs with the size of $385nm \times 238nm \times 75nm$ and a non-metallic layer $385nm \times 238nm \times 60nm$. Here, the metal is silver owning a Drude-type electric permittivity $\varepsilon(\omega) = 1 - \omega_p^2/(\omega^2 + i\omega\gamma)$, in which $\omega_p = 1.37 \times 10^{16} S^{-1}$ and $\gamma = 12.24 \times 10^{12} S^{-1}$ [2]. The middle layer and surrounding environment are both chosen to be the ytterbium-erbium codoped gain material: Er:Yb:YCOB, with the refractive index of 1.3. The gain waveguide system is placed on the $SiO_2$ substrate with index of 1.5. In such a nanosandwich waveguide, the collective magnetic resonance, MP mode, can be excited by a near-field source placed at the input of the waveguide. The energy density distribution of such MP mode with the wavelength of 1550nm is shown in Fig. 1(b). The nanosandwich waveguide can also sustain higher order modes. The energy density distribution of the high order mode in the light cone of the waveguide with the wavelength of 980nm is presented in Fig. 1(c), which is exited by a plane wave source incident on the entire waveguide plotted in Fig. 1(a). Since the nanoandwich waveguide can be considered as a chain of coupled resonators (see Fig. 1(d)), energy propagating along the waveguide is described by the following propagating equation.

$$\frac{\partial N_i}{\partial t} = \frac{N_{i-1}}{\tau_{Prop}} - 2\frac{N_i}{\tau_{Prop}} + \frac{N_{i+1}}{\tau_{Prop}} - \frac{N_i}{\tau_{Loss}} = 0 \tag{1}$$

Here, $N_i$ denotes the number of photons of signal in $i$-th nanosandwich, $\tau_{Prop}$ and $\tau_{Loss}$ correspond to the propagating and loss processes, respectively. In the steady-state case, this equation equals to zero.

When the waveguide is combined with the gain material, this equation will change and the term of gain effect has to be added into it. In this paper, we choose the gain material to be Er:Yb:YCOB, whose energy level diagram is shown in Fig. 2. In steady-state conditions, neglecting the populations in the levels $^4I_{11/2}$, $^4I_{9/2}$, and $^4F_{9/2}$ and corresponding back-transfer processes due to the fast non-radiative decay in these levels, the simplified rate equations are expressed as [12-15]

$$\frac{\partial N_{2Y}}{\partial t} = \sigma_Y v_p F_p N_p f_p (N_{1Y} - N_{2Y}) - k_1 N_{2Y} N_{1E} - k_2 N_{2Y} N_{2E} - \frac{N_{2Y}}{\tau_{2Y}} = 0 \tag{2}$$

$$\frac{\partial N_{2E}}{\partial t} = k_1 N_{2Y} N_{1E} - \sigma_E v_s F_s N_s f_s (N_{2E} - N_{1E}) - \frac{N_{2E}}{\tau_{2E}} - 2CN_{2E}^2 = 0 \tag{3}$$

$$\frac{\partial N_s}{\partial t} = \sigma_E v_s F_s \int (N_{2E} - N_{1E}) N_s f_s dV - \frac{N_s}{\tau_s} = 0 \tag{4}$$

Here, $N_{ix}$ and $\tau_{ix}$ respectively represent the population density and lifetime of the corresponding levels shown in Fig. 2. $k_1=k_2=5.0\times10^{-21} m^3/s$, are the coefficients of the two energy transfer processes. $C$, the up-conversion rate, equals to $1.3\times10^{-23} m^3/s$. $v_p$, $N_p$, and $f_p$ are the group velocity, total photon number, and normalized spatial intensity distributions of the pump light (980 nm), respectively. $v_s$, $N_s$, and $f_s$ represent the corresponding parameters of the signal light (1550nm). $f_p$ and $f_s$ are normalized as $\int f_p dV = 1$ and $\int f_s dV = 1$, where $V$ is the volume. Additionally, in steady-state conditions, it is reasonable to give the approximate expressions $N_{1E} + N_{2E} \approx N_E$, and $N_{1Y} \approx N_Y$. In our calculation, the values of $\sigma_E$, $\sigma_Y$, $\tau_{2E}$, $\tau_{2Y}$ are fixed as $5.0\times10^{-25} m^2$, $8.0×10^{-25}m^2$, $5.0×10^{-3}s$, $2.6×10^{-3}s$, respectively [11-15]. The Purcell factor is defined as $F = 3Q\lambda^3/(4\pi^2 V_m n^3)$ [16, 17], where $n$ and $\lambda$ are the refractive index of gain material and wavelength. The quality factor $Q$ and the effective mode volume of laser mode $V_m$ are respectively determined by the decay time of the mode and the field confinement. Both of them can be calculated in our simulation. Here, the coefficient $3\lambda^3/(4\pi^2 V_m n^3)$ for signal and pump light are calculated as 1/40 and 1/20, respectively. The group velocities of pump light and signal light are also calculated as $1.0×10^8 m/s$ and $0.5×10^8 m/s$.

After considering the gain effect, the propagating equation can be modified as

$$\frac{\partial N_s^i}{\partial t} = \frac{N_s^{i-1}}{\tau_{Prop}} - 2\frac{N_s^i}{\tau_{Prop}} + \frac{N_s^{i+1}}{\tau_{Prop}} - \frac{N_s^i}{\tau_{Loss}} + \sigma_E v_s F_s \int (N_{2E} - N_{1E}) N_s^i f_s dV = 0 \quad (5)$$

In the steady-state case, this equation equals to zero. Here, we choose the MP mode to be the signal light, and the higher order mode as the pump light, which leads to the larger efficiencies of pumping and radiation [11]. In general, the $Yb^{3+}$ concentration is an order of magnitude higher than $Er^{3+}$ concentration. In our calculations, we fix the $Yb^{3+}$ concentration at $5.0×10^{27}$ ions/$m^3$ [11]. The pumping power on single nanosandwich is also fixed to 0.05mW. Different $Er^{3+}$ concentration will lead to different compensation effect against the loss in waveguide. The normalized number of photons in nanosandwiches along the waveguide is shown in Fig. 3(a), with different $Er^{3+}$ concentration. Larger concentration leads to higher compensation. With $N_E$ increasing to $3.0×10^{26}$ ions/$m^3$, the propagating length enlarges to 2 times, shown in Fig. 3(a). Further increasing the doping concentration will lead to a different result, such as the case of $N_E=3.0×10^{26}$ ions/$m^3$. Although the energy of signal decreases along the waveguide, it does not go down to zero but stops at a certain value. This phenomenon is resulted from the saturation of stimulated emission radiation under certain concentration of gain ions and pumping power that can only afford a lower signal. On the other hand, we also consider the fabrication scattering and nano-size metallic structure loss by reducing the decay time of MP mode (lasing mode). We consider the energy transporting along the waveguide with different decay times of MP mode from 100Fs to 40Fs, which is the typical value in the metamaterial and plasmonic structures. The result in Fig. 3(b) shows that we can get longer propagating length with larger decay time of MP mode. It is easy to understand that lower loss leads to more evident compensation effect. And the saturation phenomenon is also found with larger decay time $\tau$=100Fs, shown in Fig. 3(b).

In the above content, it is evident that the loss is largely compensated by the gain effect especially with high $Er^{3+}$ concentration $N_E$ and long decay time $\tau$ case. In fact, the concentration of $Er^{3+}$ can be further increased to about $10^{27}$icon/m$^3$ [18], and the decay time of MP mode (signal) calculated directly from our simulation is larger than that we choose in Fig. 3 (about 110Fs). Therefore, the gain effect will defeat the loss effect and leads to the amplification phenomenon (lasing). In other word, the signal is enhanced when propagates along the waveguide, which is quite similar as a fiber amplifier. The threshold of such amplifier of nanosandwich waveguide is presented in Fig. 4 as the function of $Er^{3+}$ concentration $N_E$ and decay time $\tau$. The colors mean the ration of $N_{i+1}/N_i$, with the $N_{i+1}=1$. Therefore, the area of $N_{i+1}/N_i>1$ corresponds to the amplification case, and the points of $N_{i+1}/N_i=1$ correspond to the threshold points at different $N_E$ and $\tau$, remarking by the black dot-line curve in Fig. 4. It should be mentioned that an assumption of unvaried parameter of the gain material is made for the moderate concentration of gain ion we chose.

In summary, we introduce the Er, Yb codoped gain material to the nanosandwich waveguide in order to solve the loss problem. The loss can be largely compensated by tuning the doping concentration of $Er^{3+}$ and decay time of signal. Due to the saturation effect, the compensate cases can be divided into two types. Moreover, when the parameters exceed certain threshold, the gain effect can overcome the loss in the waveguide, then leading to the amplification of signal along the waveguide similar as the fiber amplifier. This property has a potential application in the plasmonics integrated optical circuits and metamaterials.

This work is supported by these National Programs of China (Grant Nos. 10704036, 10874081, 10534020, 60907009, and Grant No. 2006CB921804), and postdoctoral programs (20090461097 and 0204003447).

**Caption:**

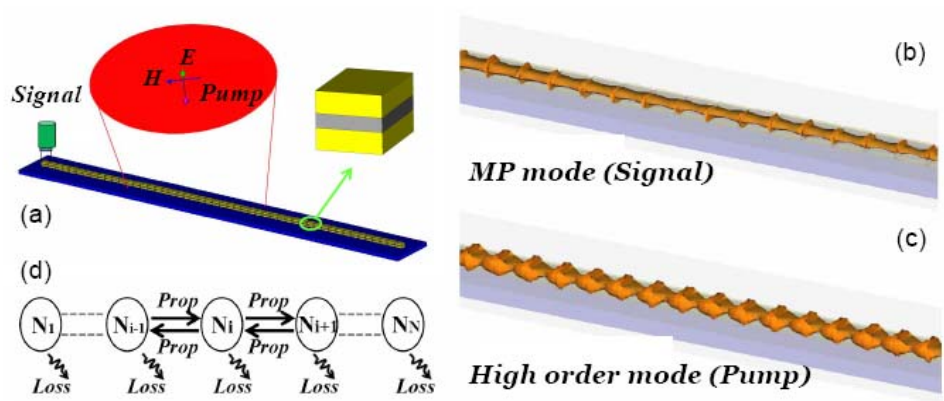

Fig. 1: The geometry of the nanosandwich waveguide is shown in (a); The MP mode and higher order mode are presented in (b) and (c), respectively; the sketch of the coupled waveguide model is plotted in (d).

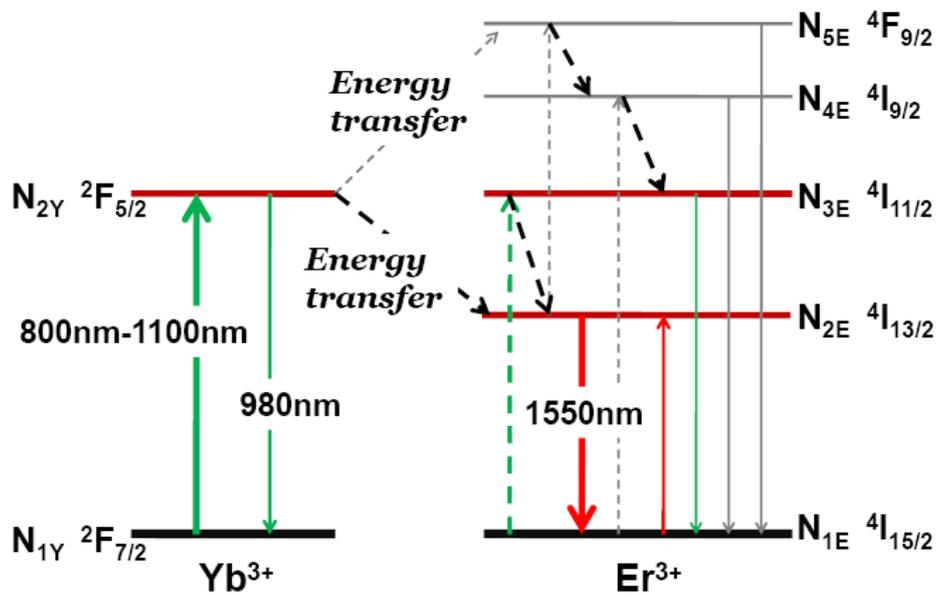

Fig. 2: The energy level diagram of the Er, Yb codoped system.

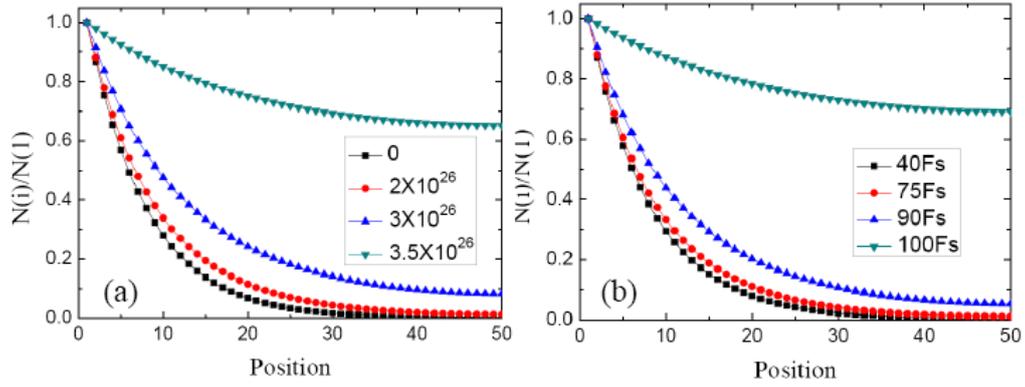

Fig. 3: The normalized numbers of photons in the nanosandwiches along the waveguide with different doping concentration $N_E$ and different decay time of signal $\tau$ are presented in (a) and (b), respectively.

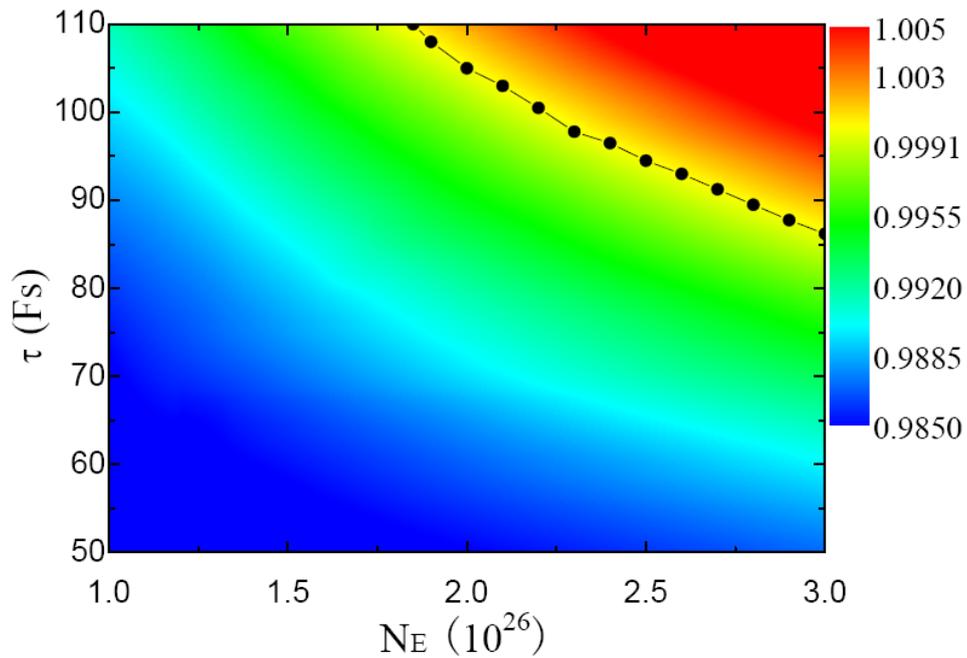

Fig. 4: The gain ability of single nanosandwich in the waveguide as the function of $Er^{3+}$ concentration $N_E$ and decay time $\tau$. The thresholds of amplification radiation are remarked by black dot-line curve.